\documentclass[a4paper,11pt]{article}
\pdfoutput=1 
\usepackage{jcappub}
\usepackage{amsmath}
\usepackage{amssymb,amsfonts}
\usepackage{graphicx,epsfig,graphics}
\usepackage{color}
\usepackage{subfig}
\usepackage{caption}
\usepackage{booktabs}
\usepackage{tikz}
\usepackage{pgfplots}
\usepackage{amssymb}
\usepackage{graphicx}

\newcommand{\m}[1]{\begin{math} {#1} \end{math}}

\newcommand{\imp}[0] { \rightarrow }

\newcommand{\MG}{\times10^{12}M_{\odot}}
\label{firstpage}
\title{Estimating the Mass of the Local Group using Machine Learning Applied to Numerical Simulations} 
\author[1]{M. McLeod}
\author[2]{N. Libeskind}
\author[1]{O. Lahav}
\author[3]{Y. Hoffman}

\affiliation[1]{Department of Physics and Astronomy, University College London, Gower Place, London WC1E 6BT, U.K.}
\affiliation[2]{Leibniz-Insitut f{\"u}r Astrophysik Potsdam (AIP), An der Sternwarte 16, D-14482 Potsdam, Germany}
\affiliation[3]{Racah Institute of Physics, Hebrew University, Jerusalem 91904, Israel}

\emailAdd{michael.mcleod.13@ucl.ac.uk}

\date{}

\abstract{We present a new approach to calculating the combined mass of the Milky Way (MW) and Andromeda (M31), which together account for the bulk of the mass of the Local Group (LG). We base our work on an ensemble of 30,190 halo pairs from the Small MultiDark simulation, assuming a $\Lambda$CDM (Cosmological Constant and Cold Dark Matter) cosmology. This is used in conjunction with machine learning methods (artificial neural networks, ANN) to investigate the relationship between the mass and selected parameters characterising the orbit and local environment of the binary. ANN are employed to take account of additional physics arising from interactions with larger structures or dynamical effects which are not analytically well understood. Results from the ANN are most successful when the velocity shear is provided, which demonstrates the flexibility of machine learning to model physical phenomena and readily incorporate new information. The resulting estimate for the Local Group mass, when shear information is included, is $4.9\MG$, with an error of $\pm0.8\MG$ from the 68\% uncertainty in observables, and a r.m.s scatter interval of $^{+1.7}_{-1.3}\MG$ estimated scatter from the differences between the model estimates and simulation masses for a testing sample of halo pairs. We also consider a recently reported large relative transverse velocity of M31 and the Milky Way, and produce an alternative mass estimate of $3.6\pm0.3^{+2.1}_{-1.3}\MG$. Although the methods used predict similar values for the most likely mass of the LG, application of ANN compared to the traditional Timing Argument reduces the scatter in the log mass by approximately half when tested on samples from the simulation. }

\begin{document}
\maketitle
\flushbottom


\section{Introduction}

Precise estimates of the mass of the Local Group (LG) remain an outstanding problem for both observers and theorists, with physical properties being difficult to observe and reliable models remaining elusive. Recent interest has been fuelled by the advent of large N-body simulations and the introduction of dark energy models. We introduce here a new approach to the problem by using machine learning to model complex phenomena present in the simulation. 

For over fifty years, the timing argument (TA), introduced in \cite{KW}, has been used as a simple dynamical estimate for the the mass of the LG, assuming that the mass is dominated by the Milky Way and Andromeda. The argument uses simple Newtonian mechanics to calculate the combined mass of the halo pair. 

The various mass estimates of the LG, some of which are discussed and compared in \cite{Raychaudhury}, demonstrate the lack of consensus in determining the LG mass. In order to build a coherent picture of the near universe, and how it fits into the universe as a whole, we require robust estimates of the LG mass and the ability to rule out particular models. Far from being an isolated system, the LG sits within a filament of the vast cosmic web, and its history is cosmology dependent through the expansion of the universe and the formation of structure. Given the dependence on \m{\Lambda} and large scale structure, we must work towards a means of utilising a variety of aspects of cosmology. A better understanding of the mass of the LG and other systems would help us to evaluate which models are successful in galaxy formation, dynamics, and near field cosmology. 

A number of more detailed models have been made to estimate the mass of the LG. \cite{Bell_Obs} includes observations on smaller bodies (other than MW and M31) in the LG. The introduction of a Cosmological Constant has also been explored (\cite{BaT}, \cite{Partridge}, and references therein), which manifests as an additional expansion term and increases the mass estimate for the LG by 13\%. The use of data from large cosmological simulations was introduced by \cite{Kroeker} and furthered by \cite{LW}, which allowed for an analysis of the robustness of the TA, and calibration of the TA mass estimate to a simulation mass to resolve biases. Li \& White find that the TA is an (almost) unbiased estimator of the mass of an LG like system, based on a selection of halo pairs selected for their similarity to the LG from the Millennium simulation, and assuming no effect from a cosmological constant. A more recent study by \cite{Gonzalez} uses simulations to calculate a likelihood for the masses, and \cite{Carlesi} uses a constrained simulation approach to generate LG like objects and study their mass distribution. 

From the previous work done estimating the LG mass, it seems clear that there is still a large uncertainty in predicting the simulation mass (\m{M_{\text{sim}}}) using available analytic models. Since these models approximate the galaxy as a point mass, and we know that galaxy and the dark matter halo which it occupies is an extended mass distribution, the mass estimate \m{M_{\text{TA}}} is not necessarily inherently physically meaningful. From a practical point of view we would like to have an estimate of more physical mass, such as the virial mass or \m{M_{200}} (the mass contained within a radius $r_{200}$ such that the density within $r_{200}$ is a factor of 200 greater than the mean density of the universe).  

In this work we combine data from N-body simulations with machine learning methods to exploit information about the environment of the LG within larger scale structure to improve estimates of the LG mass in the context of $\Lambda$CDM cosmology. We use Artificial Neural Networks (ANN), a class of machine learning algorithms which have been widely applied in other areas of physics. The network is trained by providing it with the salient physical parameters and masses for halo pairs selected from the simulations; the algorithm seeks a function which best predicts the output (combined halo mass) from the inputs (observable dynamical and local environmental parameters). Once it has converged on a solution, the relevant inputs for the LG (from observations) are fed into the function, which then returns a mass. 

In the next section we describe the simulation used, and the criteria for identifying and selecting halo pairs. In section three we review some of the TA models which have been used in the literature, apply them to our simulation data set, and look at the robustness of the assumptions of the TA. In the fourth section we give a brief introduction to artificial neural networks, and in section five we apply these techniques to mass estimation. In section six we apply our models and ANN to the case of the LG itself. 

\section{Simulations \& Selection Criteria}
\label{sims}

The simulation data is taken from the \textit{Small MultiDark Planck} (SMDPL) simulation, downloaded from the publicly available cosmosim database (www.cosmosim.org, \cite{Database}). The simulation box size is 400Mpc/$h$ ($h$ = 0.6777), with \m{3840^3} particles, with a particle mass of \m{9.63\times10^7M_{\odot}/h} and a force resolution of \m{1.5 \text{kpc}/h} (\cite{SMDPL}). Halos are identified using a friends-of-friends algorithm (\cite{Halo-Finder}). Suitable halo pairs are selected as follows:
\begin{enumerate}
	\item Candidate halos are selected with a mass \m{5\times10^{11}M_{\odot} \le M \le 10^{13}M_{\odot}}. 
	\item If the candidate is between 1.5 Mpc and 3 Mpc of a another halo of mass \m{>10^{12}M_{\odot}} then the candidate is discarded. 
	\item If the candidate is within 0.5 Mpc of a another of mass \m{>5\times10^{11}M_{\odot}} then the candidate is discarded.
	\item If there is another candidate (i.e. halo with mass \m{5\times10^{11}M_{\odot} \le M \le 10^{13}M_{\odot}}) at a distance 0.5Mpc\m{ \le r \le }1.5Mpc then the pair is accepted.
\end{enumerate}
Our criteria are less restrictive than those used in many other studies, such as \cite{LW}, because we also wish to investigate broader applicability of the TA outside of the LG. We have in total 30190 halo pairs.
This data includes the position, velocity and mass of each halo, as well as environmental data such as the local density, shear (see equation \ref{shear_def}), and velocity fields. We can easily calculate dynamical parameters \m{(r, v_r)} from \m{r = |\bf{x}_B - \bf{x}_A | } and \m{v_r = (\bf{v}_B - \bf{v}_A) \cdot \hat r }, as well as the tangential velocity \m{v_t}. 

\section{The Timing Argument and its Extensions}\label{TA_models}

In this section we will discuss some of the variants of the TA that have been proposed, and compare their performance on our dataset. These provide an important benchmark for estimators of the binary mass, as these still form a basis of our understanding of the LG mass. This is the first time different TA-like models have been systematically compared on a simulation data set, allowing us to select the best TA-based estimator. The results can be seen in Figure \ref{Contours} and section \ref{TA_Results}.

TA predicts mass as a function \m{M_{TA} = M(r, v_r, t_u)}, where \m{r} is the separation between galaxies (or halos), \m{v_r} is radial velocity, and \m{t_u} is the age of the universe (i.e. time at current epoch assuming \m{t=0} at ``big bang''). Given \m{t_u} from experiments such as Planck and the relevant cosmological model (we shall use \m{t_u = 13.8} Gyr \cite{t_u}), we may write \m{M_{TA} = M(r, v_r)}. The assumptions made by the TA are:

\begin{enumerate}

\item Galaxies have no transverse (non-radial) velocity.

\item Galaxies can be modelled as point masses.

\item Galaxy pairs are isolated; there is no external gravitational field.

\item Galaxies start their orbits in the early universe close to \m{(r,t) = (0,0)}.

\end{enumerate}

Under these simple conditions the system evolves as 
 \begin{equation} \label{TA_Lambda}
 	\frac{d^2r}{dt^2} = -\frac{GM}{r^2} + \frac{\Lambda c^2 r}{3} 
 \end{equation}
 where in most treatments (and the original formulation) one takes $\Lambda=0$. The simple form with no $\Lambda$ term has a well known cycloid solution which may be found in \cite{KW} amongst others. 

This can be extended to include transverse velocities by treating the separation and velocity as vectors, and there is an analytic solution for this case presented in \cite{Einasto}. Intuitively including tangential velocity might increase the mass estimate, but because the boundary condition is different (periapsis at $t=0$) it can in fact raise or lower the mass estimate. The results of applying this method can be compared to the traditional TA in Figure \ref{Contours}.

Another simple extension to the TA is to include a Cosmological Constant \cite{Partridge}; unfortunately, this equation no longer has a simple parametric solution and must be solved numerically. This theory result in an estimated mass increase of about 13\% for the Local Group parameters. 

This method can be extended to model other forms of Dark Energy or Modified Gravity once the appropriate radial equation is derived. These would also require a numerical integration, and would need to be compared to specially prepared non-$\Lambda$CDM simulations for statistical work (and hence will not be considered further in this paper).

\subsection{Results of applying the timing argument}\label{TA_Results}

\begin{figure}
	\subfloat[Traditional TA \label{TA_contour}]{
		\includegraphics[width=4.5cm,keepaspectratio=true]{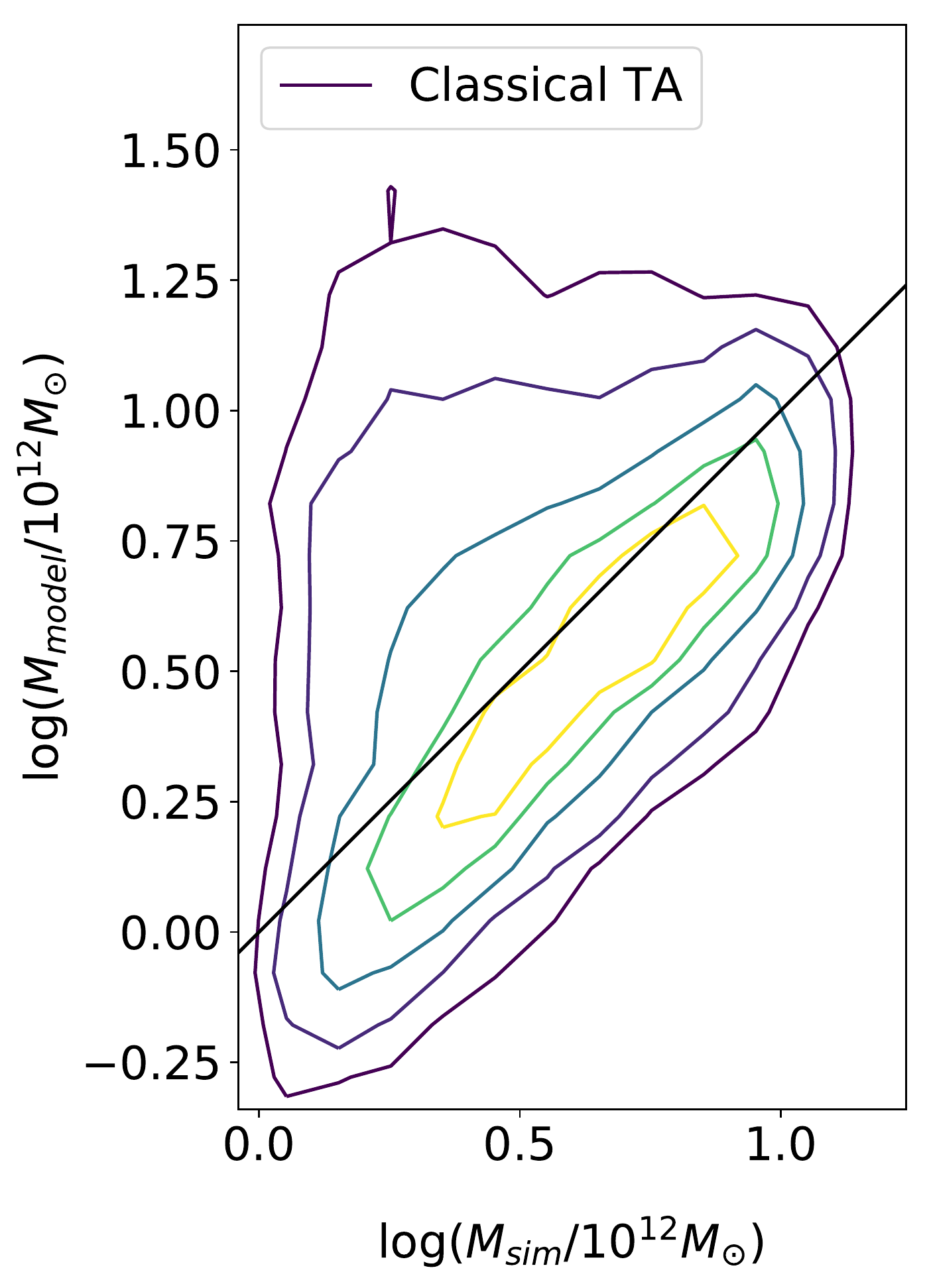} 
	}
	\subfloat[Angular Momentum \label{AngMom_contour}]{
		\includegraphics[width=4.5cm,keepaspectratio=true]{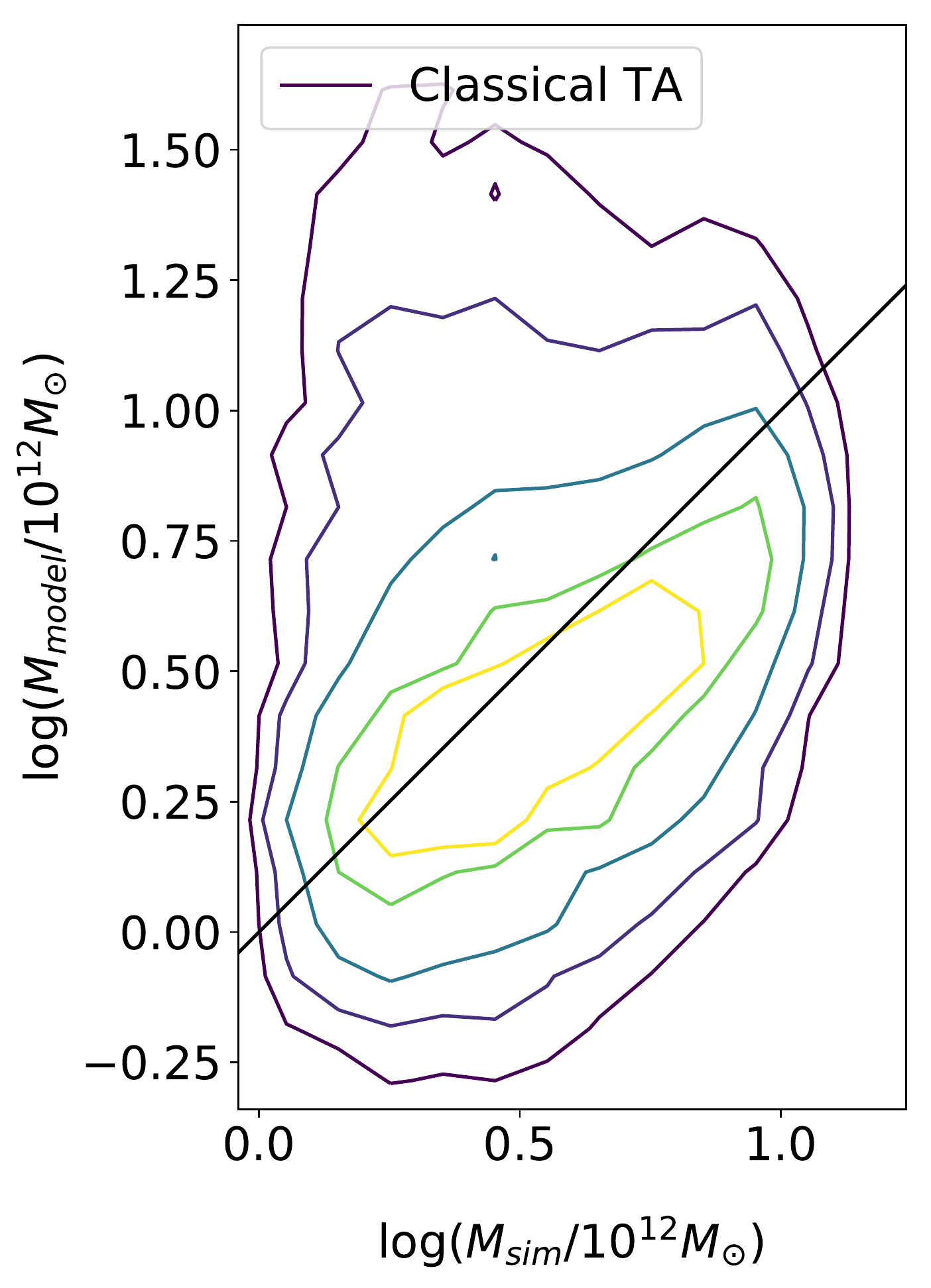} 
	}
	\subfloat[With Lambda \label{Lambda_contour}]{
		\includegraphics[width=4.5cm,keepaspectratio=true]{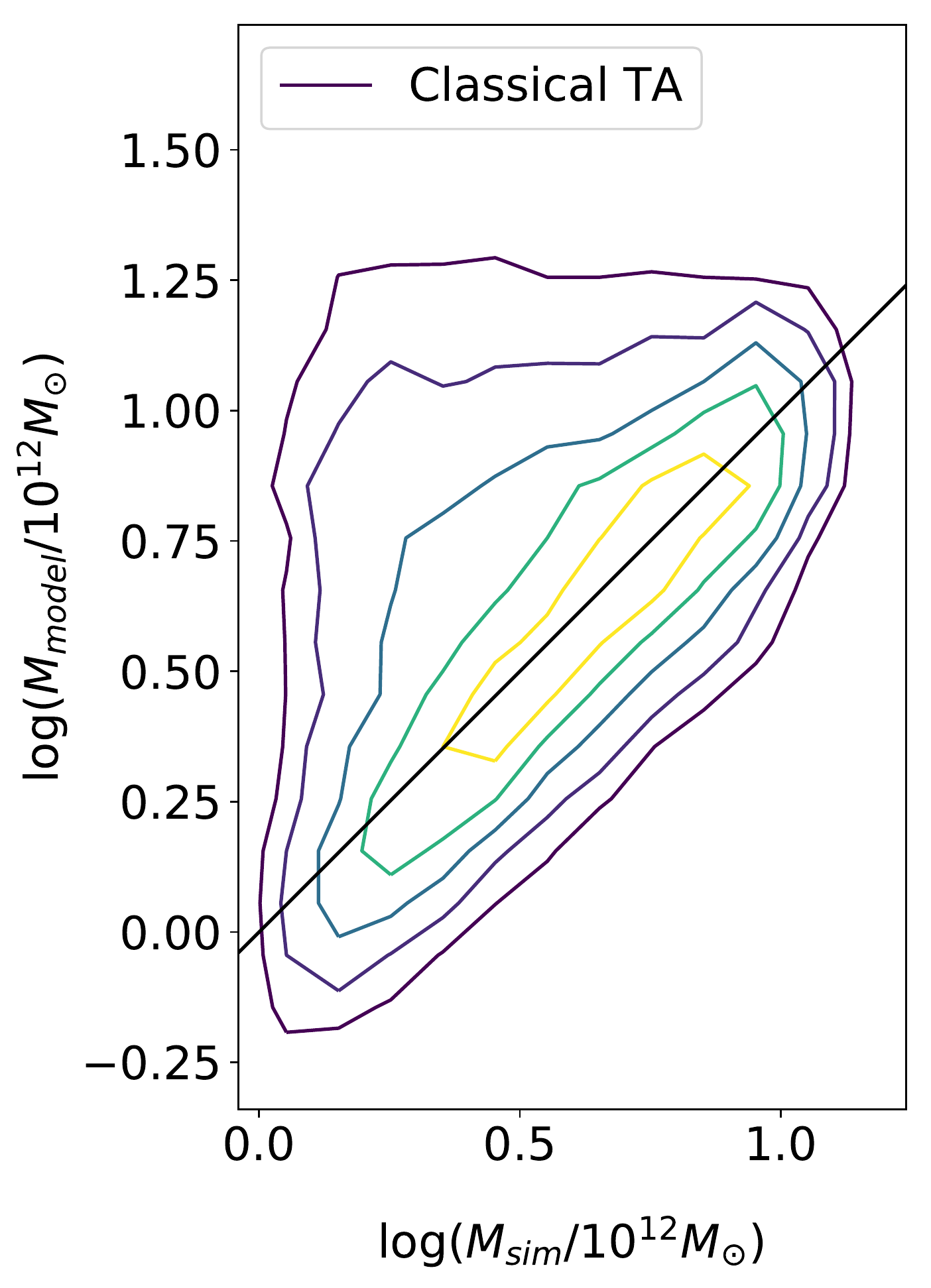} 
	}
	  \caption[.]{The results of applying the TA (left), TA with angular momentum (middle), and TA with a Cosmological Constant (right) to the data set, with the simulation mass plotted on the $x$-axis and the estimation plotted on the $y$-axis. We can see that the bias in the right panel is greatly reduced compared to the left. The middle panel shows that including $v_t$ (without $\Lambda$) does not produce an improvement, and in fact widens the contours. This may be because the tangential component of the velocity is not primordial, but acquired at later times by interaction with larger scale structure or tidal fields. Contours show the shape of a 2D histogram, with all the pairs plotted by their simulation mass and predicted mass. The contours are drawn by ordering the grid by number density of halo pairs; grid points are then added from highest density to lowest density and contours drawn at intervals where they have reached 20\%, 40\%, 60\%, 80\%, and 90\% of the pairs in the sample; over plotted is the equality line $M_{\text{sim}} = M_{\text{model}}$.}
 \label{Contours} 
\end{figure}

To obtain benchmark results for estimators of the halo mass, we may simply apply the TA models to the set of halo pairs. As we can see from Figure \ref{TA_contour}, the TA manages to match a basic trend in the data, but is unsuccessful in a wide range of low mass halos whose masses are drastically overestimated. 

The correlation between the TA estimate and the bound mass is strong in the innermost contours, although there is a visible bias when using the traditional TA (Figure \ref{TA_contour}). This bias is not remedied by the inclusion of transverse velocity to the TA (Figure \ref{AngMom_contour}). Including a Cosmological Constant appears to be effective in removing this bias (Figure \ref{Lambda_contour}). This is an important inclusion because the additional term from the Cosmological Constant models the expansion term which is present in the simulation. This highlights a dependency of the estimate on the cosmological model: for a \m{\Lambda}CDM model we must include this correction to achieve a best estimate. This may be complicated further if one wishes to consider dynamical Dark Energy models such as scalar field models or modified gravity theories.

The primary problem is that for low mass halos the TA predicts a very wide range of masses including some of the highest estimates (Figure \ref{Contours}). These account for a large fraction of the total population, with 40\% of the halo pairs being outside the three innermost contours which show a strong trend. Thus far we do not have a way of identifying a priori whether or not our galaxy lies in this region, so the uncertainty in the TA remains very large. 

We may calculate the rms scatter and Pearson product-moment correlation coefficient for this data, which provide a benchmark for further estimators which we will develop. The Pearson product-moment correlation coefficient can be calculated for a sample of data using the following equation
\begin{equation}
	p = \frac{\sum_{i=1}^{n} (x_i - \bar{x})(y_i - \bar{y})}{\left[ (\sum_{i=1}^{n}(x_i-\bar{x})^2)(\sum_{i=1}^{n}(y_i-\bar{y})^2) \right]^\frac{1}{2}}
\end{equation}

For the TA\m{_\Lambda} estimate we find that \m{p = 0.32} when we look at \m{\log(M_{TA\Lambda})} and \m{\log(M_{\text{sim}})}. The rms scatter is 0.41 in the log mass, which would correspond to a multiplicative factor of roughly 2.6 in the mass itself. 

The large scatter between the TA estimates and the simulation masses is a concern for anyone wishing to use it to estimate dynamical masses, e.g. \cite{vdMarel} and \cite{Partridge}. We are well aware of the shortcomings of the TA models, despite its many extensions; we will see if we can improve this using machine learning methods. 

\section{A Brief Introduction to Artificial Neural Networks \& Their Application to Mass Estimation}

In addition to the simple models outlined in section \ref{TA_models}, there are many complications to the history of galaxies compared to the simple Newtonian paths considered. Halos may form at different times, accumulate or shed mass in complex ways over time, or interact significantly with external tidal or gravitational fields. Rather than constructing ever more complicated analytic models from first principles, which may lead to long or unstable calculations with ambiguous boundary conditions, we instead explore the possibilities of capturing some complex behaviour by using machine learning techniques.

We wish to find a function which will calculate the mass, M, of a pair of halos based on \m{\Lambda}CDM simulations, given some input parameters; these include information about the dynamics of the system such as \m{(r, v_r, v_t)}, as well as parameters which characterise the environment such as local density or shear information. The problem is therefore a (non-linear) regression problem, in which we require an algorithm to find a best fit model to our simulation pairs from our inputs. ANN are one such class of machine learning algorithms, which model unknown functions using a combination of compositions of (typically) sigmoid functions. The particular program that we use is called \emph{ANNz}, an ANN code developed by \cite{ANNz} for the purposes of predicting photometric redshifts (although the technique is general). The composition of weighted sigmoid functions is unintuitive, and makes their output more difficult to interpret, but means that they are much more flexible with respect to the kinds of functions that they can model, and do not require any special knowledge about the problem which might be needed in order to choose an appropriate set of basis functions. In our case we do not have an obvious set of basis functions which could be used since we have no clear idea of how our additional parameters will affect our mass estimates in terms of functional forms. Thus we may resolve that ANN are a good starting point for modelling this behaviour. 

The interested reader may consult  and \cite{ANN}, \cite{ANN_book}, \cite{ANNz}, \cite{ANNz2}, (amongst many others) for background information on ANN algorithms, their theoretical properties, and their manifold applications.

\subsection{Applying ANN to mass estimation}

To apply the ANN to the problem at hand our observables such as $r$ and $v_r$, divided by some appropriate unit of measurement, are used as input values i.e. the first layer of our network. The simulation masses are the target outputs used in the training and validation sets. In order to produce more stable calculations and avoid large weights (which lead to very large regularising terms), units are scaled so that all inputs and outputs are within a few orders of magnitude of unity, with the following units being used:
\begin{enumerate}
	\item Mass in \m{M_G = 10^{12} M_{\odot}}
	\item Time in Gyr \m{= 10^9} years
	\item Distance in Mpc
	\item Velocity in Mpc Gyr\m{^{-1}}
\end{enumerate}
The network needs only one node in the final layer for the output, which is the predicted mass in $M_G$.
The ANN algorithms have randomised elements which require a random seed which can produced small variations in the results; because of this we use an ensemble of five ANN regressions with different seeds to produce each result.

All ANN runs were required to converge to be included in the ensembles. The architecture used for the results included in this paper had 5 hidden layers of 10 nodes each; more complex architectures did not yield an improvement indicating that the functions were not being limited by their number of free parameters. 

The training, validation, and testing sets include $\approx 5000$ halo pairs each, selected at random from the full set (subject to any data cuts) and with no halo pair appearing in more than one of the three sets. These contain, for each pair, the input parameters $\{r, v_r, ...\}$ and the target parameter $M_\text{sim}$. Performance is tested by comparing the predicted values of the ANN on the testing set with the genuine values $M_\text{sim}$ for each pair therein. 

\subsection{Comparing the ANN to the TA with $\Lambda$}

We would like to get an idea of how much improvement the machine learning methods have offered compared to our earlier application of the TA. We can measure then the rms scatter of the predictions from their `true' mass taken from the simulation. We apply the ANN directly to \m{\log(M_{\text{sim}})} using $(r,v_r)$ as inputs, and compare this to the prediction from the TA. 

To calculate the scatter we the use simple measure:
\begin{equation}
	s_{rms} = \sqrt{\frac{(\log(M_\text{p}) - \log(M_{\text{sim}}))^2}{N}}
\end{equation}
Where \m{M_p} is the predicted mass (ANN or TA), \m{M_{\text{sim}}} is the mass from the simulation, and \m{N} is the number of points in the sample. This is the measure that is typically taken (plus an additional regulation term) as the cost function of the ANN and other machine learning algorithms. 

\begin{figure} 
\subfloat[ANN, run with only $(r,v_r)$ as input, and TA estimates of mass against the bound mass from simulations. \label{TA_ANN_Bound}]{\includegraphics[width=7cm,keepaspectratio=true]{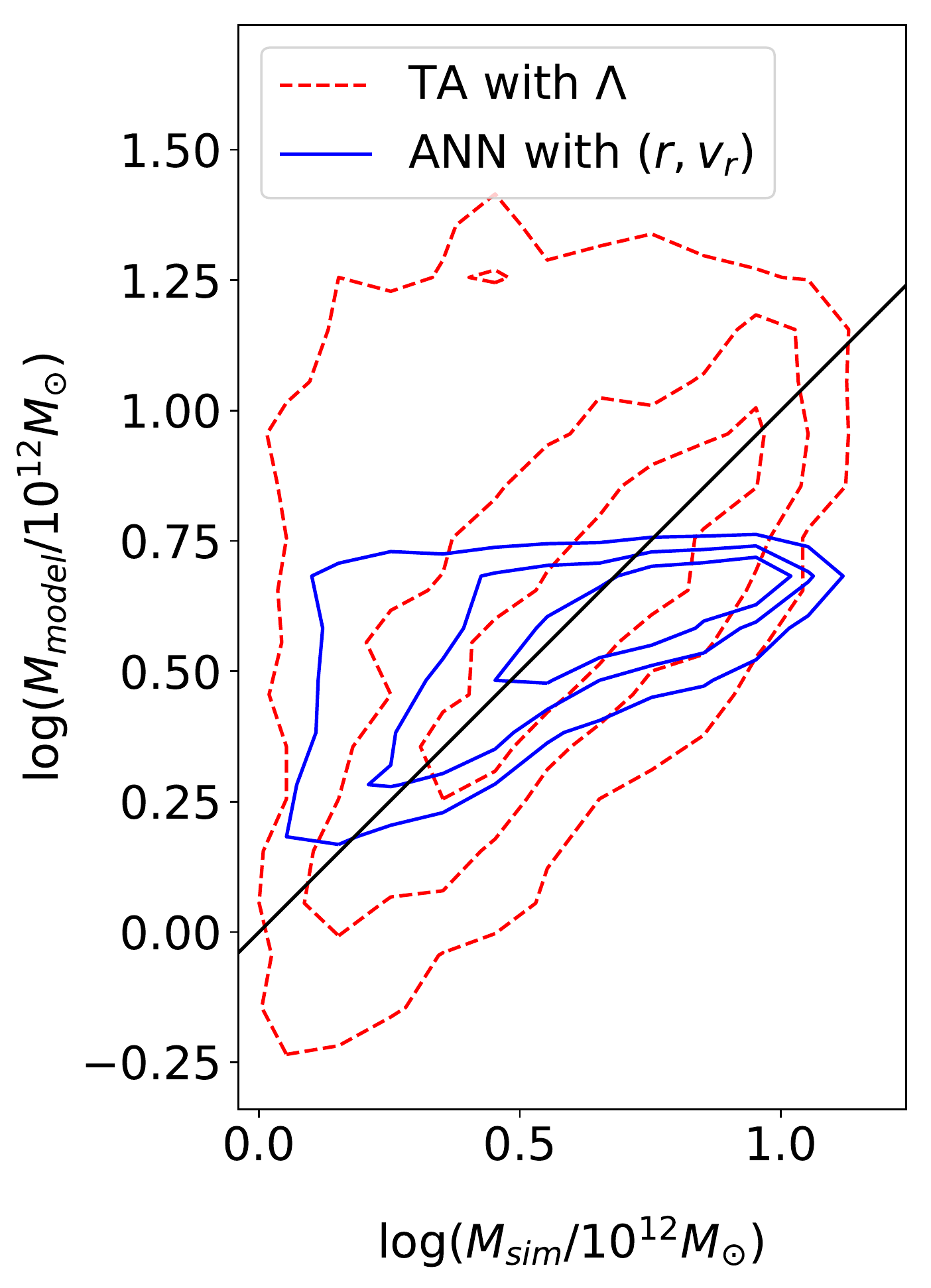}}
\hspace{1cm}
\subfloat[Contours comparing the ANN model utilising shear with the mass in the simulation. The result shows significantly better correlation than ANN estimates without shear (correlation coefficient = 0.63). \label{Contours_Shear}]{\includegraphics[width=7cm,keepaspectratio=true]{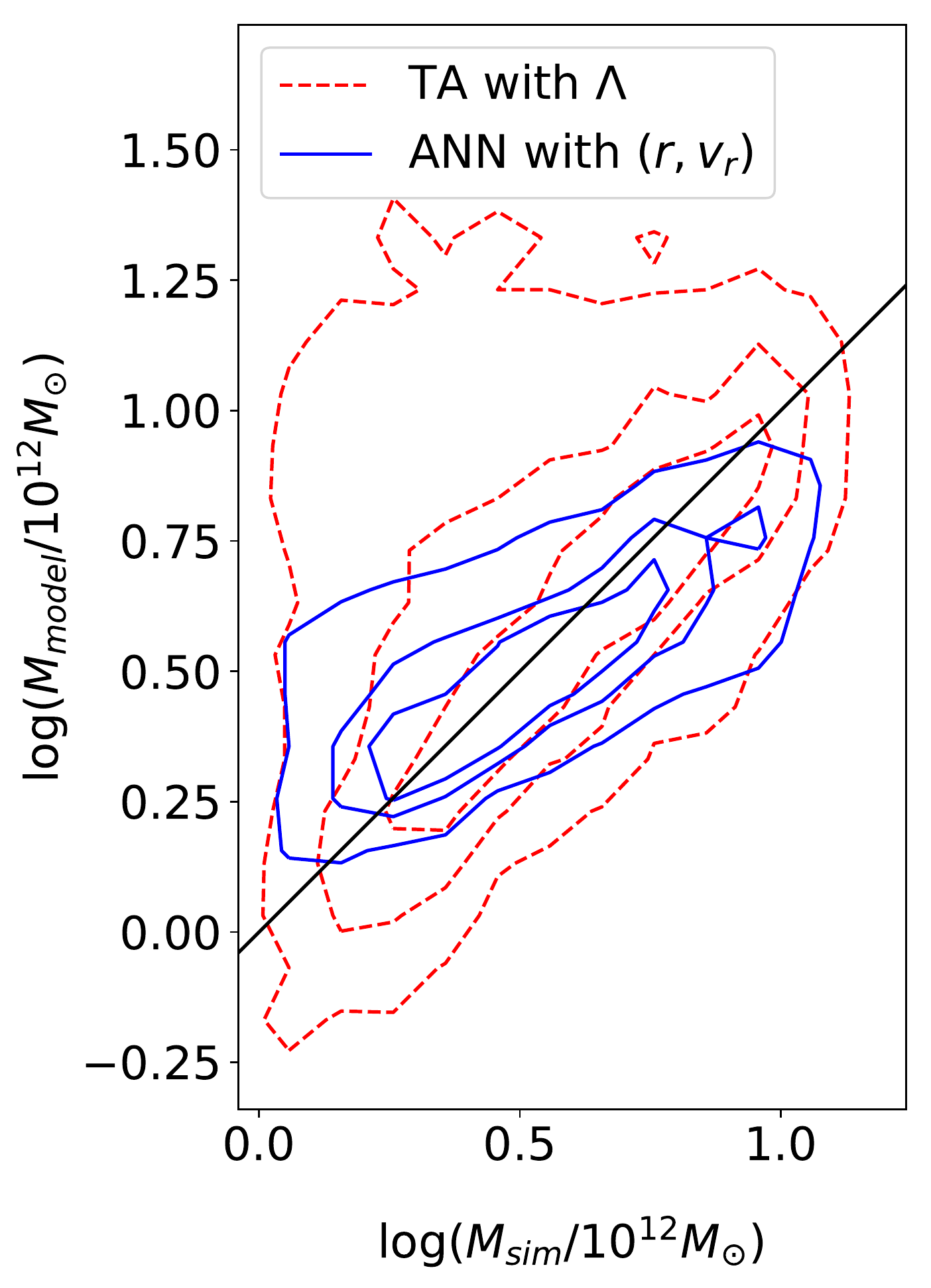}}
\caption[.]{Contours comparing the mass estimates from the ANN with those from the TA. TA (with a Cosmological Constant) is shown by the dashed red lines, the relevant ANN estimate in solid blue, with the diagonal (equality line) overplotted. Points here are taken from the `testing' subset of the data (roughly 5000 points); contours are drawn as in Fig. 1, but due to the smaller sample size are rougher and enclose 30\%, 60\%, and 90\% of the pairs. The equality line is over plotted in black.}
\end{figure}

The results are summarised in Table \ref{table:ANN_vs_TA} and Figure \ref{TA_ANN_Bound}. At a first glance, the numerical results in Table \ref{table:ANN_vs_TA} suggest that machine learning methods can reduce the scatter by almost half using only the same inputs as the analytic model, and significantly improve the correlation coefficient as well. This would imply that if we feel confident in our distribution of halo masses within large cosmological simulations, then machine learning methods may provide a significant improvement upon TA$_\Lambda$. 

\begin{table}
\begin{tabular}{|c|cc|}  
\hline
Model & rms scatter & Pearson correlation\\ 
\hline
TA with $\Lambda$                           & 0.43  & 0.32  \\ 
\hline
ANN                    &    0.23   & 0.53   \\ 
\hline
\end{tabular}
\caption[Table caption text]{rms scatter in the log between predicted mass and simulation masses, using the TA or the ANN to predict the masses of the halo pairs using only the dynamical parameters $(r,v_r)$.}
\label{table:ANN_vs_TA}
\end{table}

Figure \ref{TA_ANN_Bound} however demonstrates an interesting problem. The shape of the contours are not well in line with the equality line $M_p = M_{\text{sim}}$, indicating that the estimate is highly biased, and the shape of the contours is notably compressed into the middle of the range. Considering that the TA is very successful for many halo pairs, and that it is easily reconstructable by an ANN, can we understand why the ANN produced the behaviour that it did? Despite the fact that the TA appears to show much less bias for many halos, there are a very large number of low mass galaxies which are drastically overestimated by the TA. These would contribute very heavily to the cost function of the ANN $\sim \sum (M_\text{p} - M_\text{sim})^2$ if it tried to reproduce the TA. With only $(r,v_r)$ as inputs however, the ANN cannot find a solution which is satisfactory for these outlier halos, and therefore finds it beneficial to compress the estimates down into a narrower mass region clustered around the mean mass. This has happened to an extent that the function appears to have little visible correlation with the mass; high mass estimates in particular (where the simulation mass spans almost the entire range) are moved towards the centre of the mass range to minimise the scatter. This is a fairly generic problem in machine learning when presented insufficient information to solve a problem: the best way to reduce the scatter is to cluster around the mean. If the TA is to be improved upon, we will need to find out why some galaxies have much lower masses than we na\"ively expect, or find a parameter which can break the degeneracy. 

\section{Extending the Model and Cuts to the Data}

We now look at the effect of using some of the additional parameters we have at our disposal from the large simulation data set. 

\begin{table}
\centering
\caption{Table showing the rms scatter in the log mass for each ANN model using different input parameters and cuts to transverse velocity, alongside the TA$_\Lambda$ for comparison. For all ranges of $v_t$ the ANN with input $(r,v_r,\lambda_i,\mu_i)$ reduces the scatter in the log significantly compared to the TA: by approximately 50\% when there is no cut on $v_t$, or approximately 40\% when there are strict cuts.}
\label{scatter_table}
\begin{tabular}{|c|c|c|c|c|c|}
\hline
Max $v_t$ & $(r,v_r)$ & $(r,v_r,v_t)$ & $(r,v_r,\rho)$ & $(r,v_r,\lambda_i,\mu_i)$ & TA$_\Lambda$ \\ \hline
None     & 0.23  & 0.23                & 0.23    & 0.21  & 0.41 \\ \hline
500 km/s     & 0.23  & 0.23                & 0.23    & 0.20 & 0.39 \\ \hline
250 km/s    & 0.23  & 0.22                & 0.22    & 0.20 & 0.35 \\ \hline
125 km/s   & 0.20  & 0.20                & 0.20    & 0.17 & 0.28 \\ \hline
62.5 km/s  & 0.18  & 0.19                & 0.17    & 0.13 & 0.21 \\ \hline
\end{tabular}
\end{table}

\subsection{Transverse velocities} 

The first additional parameter used was the transverse velocity. We use the modulus of the transverse velocity as a one dimensional quantity rather than using a two dimensional vector, since the direction (which must be perpendicular to the radial vector) should not be relevant in the absence of other information. Thus \m{v_t = |\bf{v} - \bf{v}_r|}. We run the ANN with input \m{(r, v_r, v_t) \rightarrow \log(M_{\text{sim}})}.

Adding the transverse velocity on its own has little effect on the scatter (see Table \ref{scatter_table}), where we show the scatter in the log for ANN results with different additional parameters and cuts. 

\subsection{Cuts to transverse velocity}

Cuts to the transverse velocity have a significant impact on the scatter (see Table \ref{scatter_table}). Samples which are limited to small transverse velocities show significant improvement in both the TA and in the ANN. For the TA this is obvious -- the argument is based upon the assumption of a radial orbit. The ANN improves alongside it, with the ANN scatter remaining $\sim 40 - 50\%$ less than the scatter for the TA for each cut. This may be because the range of possible orbits that the ANN has to reconstruct is restricted when the system is closer to one dimensional, leading to a simpler target function. 

\subsection{Environmental parameters}

We may be able to improve our results by including parameters which characterise the environment of the halos. We are interested in observable parameters which may affect the dynamics of the pair, such as the large scale density field, velocity and shear fields. 

Environmental parameters show that some improvement may be made compared to the application using only the traditional dynamical variables (distances, velocities). 

\subsubsection{Density}

The density parameter \m{{\rho}/{\bar{\rho}}} (smoothed over 2Mpc) was added to the ANN, although the improvement is not significant (see Table \ref{scatter_table}). Perhaps a more fruitful parameter would be a density gradient vector, which would give an indication of the strength and direction of the external gravitational field which may perturb the orbit, although this is beyond the scope of this paper. 

\subsubsection{The velocity shear tensor}
\begin{figure} 
  \caption{Error histograms for the TA and the ANN run with shear information (with no cuts on $v_t$). The ANN manages to avoid the large tail of overestimates that is produced by the ANN. $\Delta \log(M) = \log(M_\text{sim}) - \log(M_\text{model})$.}
   \label{LG_errors}
\includegraphics[width=8cm,keepaspectratio=true]{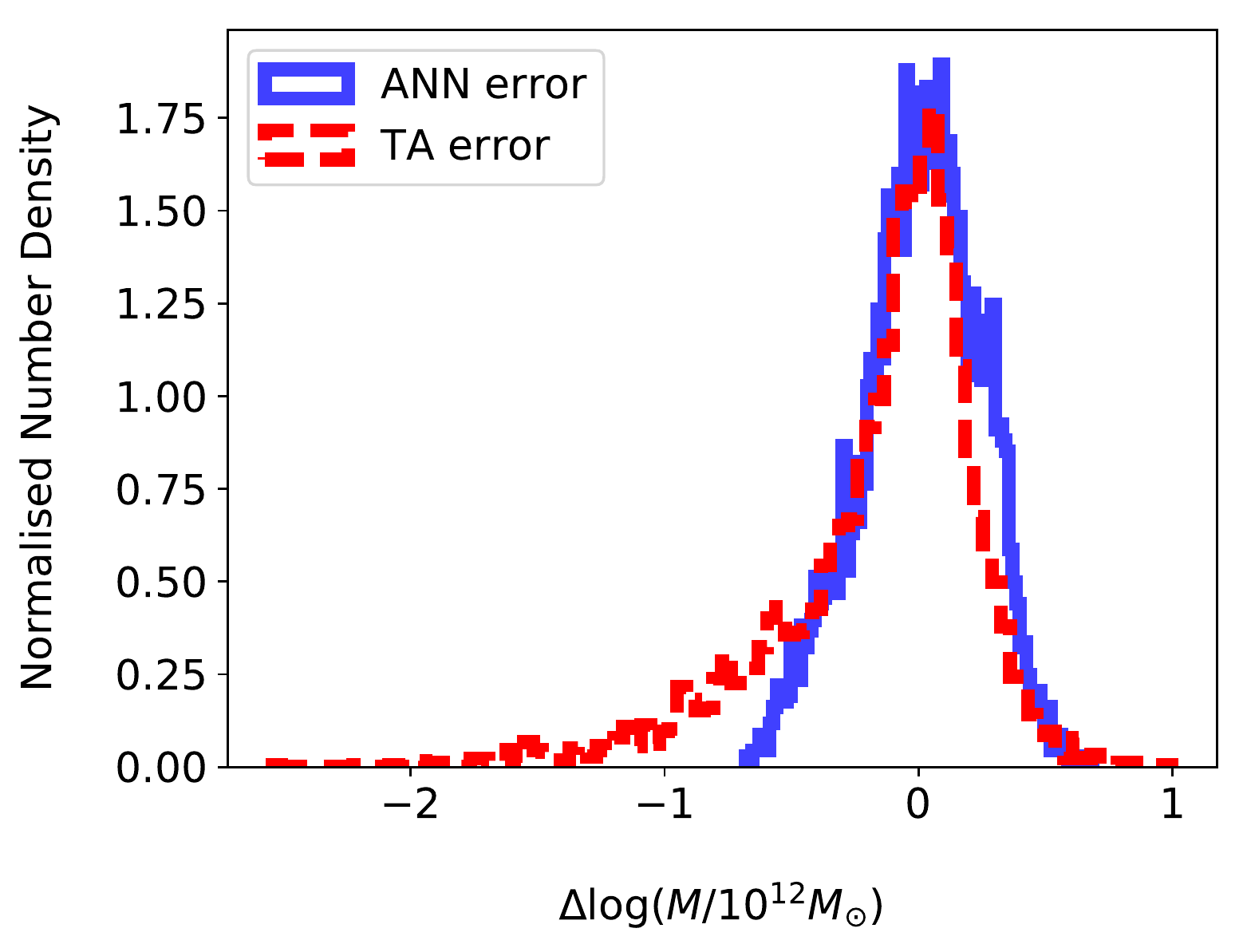}
\end{figure}

The most useful parameter investigated in this paper is the shear tensor (also smoothed over 2Mpc). The shear tensor (\cite{Shear}, \cite{Hoffman}) is defined as 
\begin{equation}
	\label{shear_def}
	\Sigma_{ij} = - \frac{1}{2H_0} \left( \frac{\partial v_i}{\partial r_j} + \frac{\partial v_j}{\partial r_i} \right)
\end{equation}
At a given point the eigenvalues and eigenvectors of this tensor can be calculated. The shear tensor has three eigenvalues \m{\lambda_{1,2,3}} with three unit eigenvectors \m{\hat{e}_{1,2,3}}. We order the eigenvalues such that:
\begin{equation}
\lambda_3 \le \lambda_2 \le \lambda_1
\end{equation}
We calculate the absolute value of the cosine of the angle between the eigenvectors and the radial separation vector \m{\mu_i = |\cos(\theta_i)| = |\hat{e}_i \cdot \hat{r}|}. (We must use the absolute value since the eigenvector only defines an axis rather than a specific direction, so alignment with \m{\hat{e}_i} or \m{-\hat{e_i}} is physically identical.) We use the ANN with an input of the form \m{(r,v_r,\lambda_1,\mu_1,\lambda_2,\mu_2,\lambda_3,\mu_3) \rightarrow \log(M)}. This provides the ANN with information about how the orientation of our physical system lines up with the principles axes of any local expansion or collapse. The scatter in the log mass, as shown in Table \ref{scatter_table}, reduces by about 10--15\% compared to the most basic form using \m{(r,v_r)} with an equivalent cut on \m{v_t}. Perhaps more importantly, there is a change in shape of the contours between Figure \ref{TA_ANN_Bound} (the results without environmental information) and Figure \ref{Contours_Shear} (the results utilising shear information). The upper bound on the estimated masses has now been relaxed and the results look significantly more intuitive, with a less dramatic bias. We can see from Figure \ref{LG_errors} that the tail TA$_\Lambda$ overestimates is significantly reduced. 

Why might the inclusion of shear have improved things? The velocity shear tensor contains information about the movements of matter on large scales, and hence may be linked to variations in behaviour from what one would expect in an isolated environment. The tensor may be used as a map of the cosmic web (as in \cite{Shear}, \cite{Hoffman}), with the sign and magnitude of eigenvalues determining if a halo is occupying a knot, filament, sheet, or void. Such a classification method has already been used to examine how halos and galaxies align with the large scale structure (\cite{tracer}, Forero-Romero, Contreras, \& Padilla 2014, \cite{Pahwa}), how galaxy properties are affected by their cosmic web environment (\cite{Nuza}, \cite{Metuki}), and how LG-like systems are formed in filaments (\cite{Forero}). 

\begin{figure}
\subfloat[The dependence of ANN mass estimate on the parameter $\lambda_2$ using $\mu_2=1$ (i.e. the system is totally aligned with the second eigenvector $e_2$), $\mu_{2,3}=0$. To respect the ordering of $\lambda_{1,2,3}$ that we have used, the upper and lower eigenvalues are set at $\lambda_1 = 0.15$ and $\lambda_3 = -0.15$. Negative $\lambda$ implies expansion and positive $\lambda$ implies collapse along the direction $\bf{\hat{e}_i}$. Masses estimates are higher in expanding regions and lower in collapsing regions. \label{lgraph}] {\includegraphics[width=7cm,keepaspectratio=true]{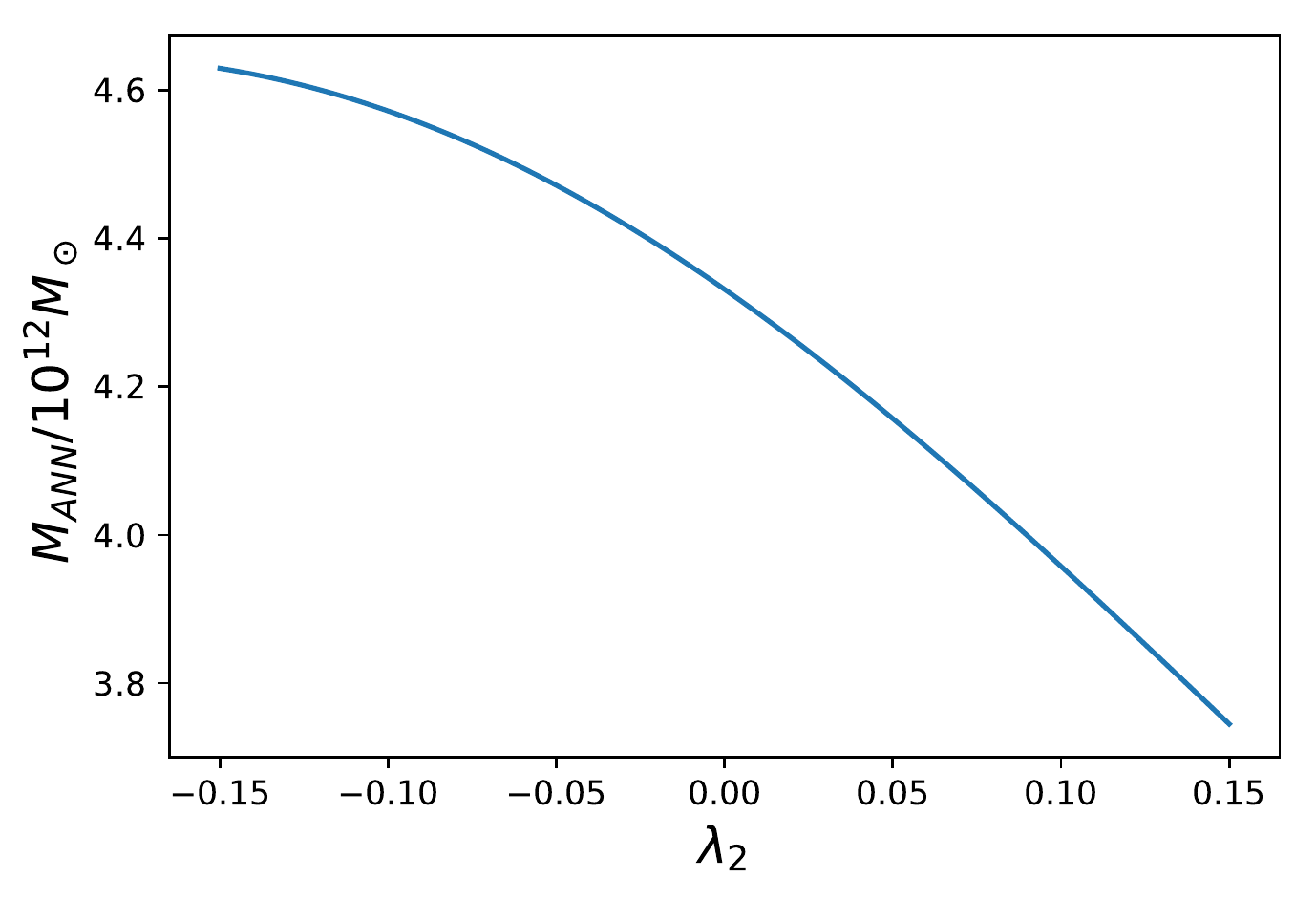}}
\hspace{1cm}
\subfloat[The dependence of the ANN mass estimate on the alignment with the eigenvectors. Here we have held the eigenvalues at $\lambda_1 = 0.15$, $\lambda_2 = 0$, and $\lambda_3 = -0.15$ i.e. collapse along the $e_1$ direction, no shear on the $e_2$ direction, and expansion along the $e_3$ direction. The angle is measured between the radial separation vector and the eigenvector $e_2$, when rotating in the $e_3 \times e_2$ plane (from full alignment with $e_2$ to full alignment with $e_3$), and rotating in the $e_1 \times e_2$ plane (from full alignment with $e_2$ to full alignment with $e_1$). We see that as we rotate to align with the collapsing vector ($e_1$) our mass estimate decreases, and as we rotate to align with the expanding vector ($e_3$) our mass estimate increases. \label{cgraph}]{\includegraphics[width=7cm,keepaspectratio=true]{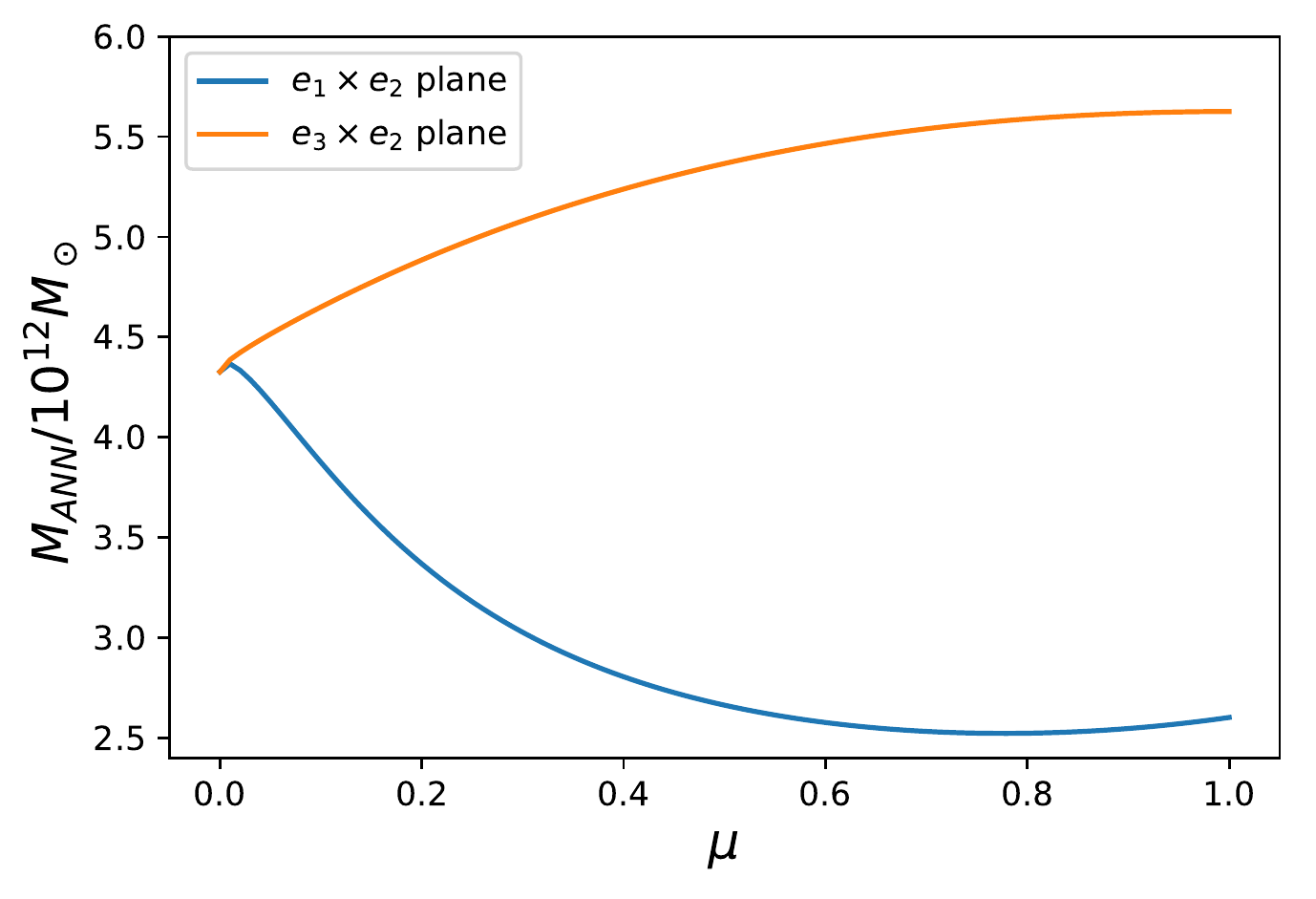}}
\caption[.]{Cross sections of the ANN function $F_{ANN} : (r,v_r,\lambda_i,\mu_i) \imp \log(M)$ where one parameter is varied and the others kept constant to visualise the mass dependence on that parameter for a LG-like halo pair. In both cases $r = 0.77\text{ Mpc}$, and $v_r = -130\text{ km s}^{-1}$, and these parameters are not varied.}\label{ShearPlots}
\end{figure}

In order to understand something about the way that the shear is influencing the solution, we look at the output with some physical insight and intuition. Whilst the fundamental representation of the function -- encoded in dozens of weights -- is no doubt difficult to interpret on its own (it is, after all, a rather garbled way to write a function and even the most brilliant mathematician would balk at diving into a quagmire of sigmoid functions analytically), we may achieve significant insight by plotting cross sections of this multidimensional function. By varying just one variable at a time and holding all others constant we may see the impact of that variable on systems which are otherwise totally similar. For example, we may hold $(r,v_r)$ fixed, thus considering only systems which have identical observable dynamics at present, and vary just one of the shear magnitudes $\lambda_2$. (We choose $\lambda_2$ since we must always respect $\lambda_3 \le \lambda_2 \le \lambda_1$ which can be done by setting $\lambda_{1,3}$ and varying $\lambda_2$ between them.) This is shown in Figure \ref{lgraph}. We wish to inspect a case where $\lambda_2$ has great impact and the other shear parameters have as little as possible, so we choose a system which is aligned totally with $e_2$. The result is the mass estimate decreases with increasing $\lambda_2$, i.e. with increasing degree of `collapse' along the radial axis of the system. This is the behaviour of the function that has been derived, and now it is prudent to ask whether this makes physical sense. In order to ask such a question, we must be very sure about what has changed and what has not. The background has changed, but the observed dynamics have not. Let's compare to a case with no shearing at all -- two bodies orbit one another in empty space. Heavy halos turn around sooner and will fall into one another faster; lighter halos will turn around later and fall into one another more slowly. Let's say in the empty space scenario, a mass of $M_{*}$ will yield the dynamics we see $(r,v_r)$. Now let us consider the case where there is a larger scale tendency towards compression along the radial axis (high $\lambda_2$). Our smaller bodies are also part of this flow, and will be involved in this larger scale collapse. This will bring us to a quicker turnaround than the empty space case, and thus we will not produce the same dynamics with $M_{*}$. In order to delay our turnaround and slow the collapse back down to our observed dynamics, we must reduce the mass estimate, and so we have a collapsing mass $M_{C} < M_{*}$. The same reasoning may be applied to the expanding case to yield an expanding mass $M_{E} > M_{*}$. 

It is worth at this point addressing some issues of structure formation. We may look towards the Raychaudhuri equation, or its Newtonian equivalent, as a means of calculating the mass in the presence of shear. Here, crucially, the shear only enters as $\sigma^2 = \sigma_{ij}\sigma^{ij}$, and thus any shear regardless of sign or alignment (the case is spherical) will enter in the same way; when applied to structure formation we obtain a reduced mass. As a result, one might expect that we should see our mass estimate reduce in the presence of any shearing, but this is not the case. Remember that we are not tracking the growth of an initial over density, but looking at the dynamics of two bodies and asserting the mass which would produce it. The presence of shear might cause the galaxies to initially form with lower masses than they would otherwise, but the mass with which they are formed is irrelevant to our purposes: we only see deviations of the trajectories from empty space dynamics, e.g. when light galaxies look heavier than they are due to their distorted dynamics. This is very similar to why the density parameter is not useful -- it affects the formation of halos but, if it is not much varying over the distance scale of the orbit, it does not affect much the dynamics of the halos.

Figure \ref{cgraph} explores the significance of the alignment of the radial vector with the shear eigenvectors. The eigenvalues are set up so that one axis is `collapsing', one axis is `expanding', and the other is `neutral' ($\lambda = 0$). The observed dynamics $(r,v_r)$ are again kept constant, and the alignment of the radial vector is rotated (via the parameters $\zeta_{1,2,3}$) from total alignment along one eigenvector to the next. The results show that the impact of the collapsing and expanding effects that we previously discusses are dependent on the alignment of these axes with the radial axis. If there is collapse close to orthogonal to the radial system, it does not much impact the trajectories along the radial line, so there is little change to mass. This is physically to be expected. 


\subsubsection{Shortcomings of additional parameters}
The scatter is still very significant, and the additional parameters make a relatively small impact on the result. Our inability to reduce it further may be for a number of reasons
\begin{enumerate}
	\item There are further important variables which may have a more significant effect. 
	\item The relationship between the dynamics of the pair and the environment is highly complex and non-linear, perhaps even chaotic.
	\item We only have a snapshot of these parameters at the present time; they are however dynamical and the impact on dynamics would depend on the variation in these parameters throughout the history of the universe.
\end{enumerate}

\section{Application to the Local Group}

\begin{table}
\centering
\caption{A table of observed velocity values for the LG from three papers. van der Marel \& Guhathakurta (2012) is the most reliable recent result. van der Marel \& Guhathakurta (2008) is included for comparison with earlier papers such as Partridge et al. 2015, which use values for the LG closer to this. Salomon et al. (2015) is a controversial recent result suggesting very different relative motion of MW and M31, which is included to explore the implications of such a result. These parameters will be referred to in the text and table as vdM. 2008, vdM. 2012, and Sal. 2015 respectively.}
\label{vdM_Sal}
\begin{tabular}{lllll}
\cline{2-4}
\multicolumn{1}{l|}{}       & \multicolumn{1}{l|}{vdM. 2008}    & \multicolumn{1}{l|}{vdM. 2012}        & \multicolumn{1}{l|}{Sal. 2015}        &  \\ \cline{1-4}
\multicolumn{1}{|l|}{$v_r$ / km s$^{-1}$} & \multicolumn{1}{l|}{$-130 \pm 8$} & \multicolumn{1}{l|}{$-109.4 \pm 4.4$} & \multicolumn{1}{l|}{$-87.5 \pm 13.8$} &  \\ \cline{1-4}
\multicolumn{1}{|l|}{$v_t$ / km s$^{-1}$} & \multicolumn{1}{l|}{$42\pm56$}    & \multicolumn{1}{l|}{$17\pm17$}        & \multicolumn{1}{l|}{$149.4\pm92.3$}   &  \\ \cline{1-4}
                            &                                   &                                       &                                       & 
\end{tabular}
\end{table}

\begin{table}
\centering
\caption{A table presenting the mass estimates for the local group for each model, as applied to the input parameters with velocities from vdM. 2008, vdM. 2012, and Sal. 2015 (see Table 3). Quoted errors are first for propagating the errors in the observables through the TA/ANN functions, and second for $\pm$ the r.m.s. scatter on the log mass between the TA/ANN model and the simulation set. For vdM. 2008 and vdM. 2012 we may use strict cuts on $v_t$, whereas for Sal. 2015 we must use cuts that include the higher transverse velocity, thus leading to a higher uncertainty. The Bayesian results quote only a 68\% confidence interval.}
\label{LG_table}
\renewcommand{\arraystretch}{1.5}
\begin{tabular}{|c|c|c|c|}
\hline
& \multicolumn{3}{c|}{$M_{LG}$ / $10^{12}M_{\odot}$} \\
Model & (vdM. 2008)  & (vdM. 2012) & (Sal. 2015)  \\
\hline 
TA$_{\Lambda}$   & $5.8^{+1.0 +3.6}_{-0.9 -2.2}$ &    $4.7^{+0.7+2.9}_{-0.6-1.8}$       &      $3.8^{+1.1+4.7}_{-0.9-2.1}$ \\
\hline 
ANN    & $3.7^{+0.3+1.9}_{-0.3-1.3}$   &    $3.6^{+0.3+1.8}_{-0.3-1.2}$         &      $3.3^{+0.6+2.3}_{-0.5-1.4}$   \\
\hline 
ANN + Shear & $6.1^{+1.1+2.1}_{-1.1-1.6}$     & $\mathbf{4.9^{+0.8+1.7}_{-0.8-1.3}}$       &       $3.6^{+1.3+2.1}_{-1.1-1.3}$ \\
\hline
Bayesian  &  $3.4^{+1.9}_{-1.2}$  & $3.1^{+1.3}_{-1.0}$       &       $3.4^{+2.3}_{-1.3}$    \\
\hline
\end{tabular}
\end{table}

We may apply what we have learned directly to the LG i.e. the Milky Way (MW) and its most massive partner, Andromeda (M31). We have learned that the TA's range of applicability may be narrow, and both the TA and ANN are more successful if we make cuts to look at LG-like galaxy pairs. Results of the different methods of calculating the LG mass is summarised in Table \ref{LG_table}.


\subsection{Input parameters}
We use \m{r = 770 \pm 40} Mpc, and velocity parameters from three studies referred to as vdM. 2008, vdM. 2012, and Sal. 2015 (see Table 3 for details). The most reliable case is represented by vdM. 2012, from \cite{vdMarel}. vdM. 2008 is taken from an earlier paper and included only in Table \ref{LG_table} for comparison to earlier results. \cite{Transverse} (Sal. 2015) represents a controversial new study which suggests a non traditional relative motion, which a much larger transverse motion. For both cases we will use the same range of \m{r}, as it contains the values of \m{r} used in both papers. 

The shear parameters of the local universe are taken from \cite{LG_Shear}, and are \m{\lambda_1 = 0.148\pm0.038}, \m{\lambda_2 = 0.051\pm0.039}, and \m{\lambda_3 = -0.160\pm0.033}. To calculate the angle between the eigenvectors and the MW-M31 separation vector, we use \cite{LG_Shear} for the shear eigenvectors and \cite{M31_pos} for the direction of the separation vector. 

\subsection{TA estimates}
Our most accurate TA model is the TA with \m{\Lambda}, which applied to the LG with vdM. 2012 gives a mass estimate of \m{4.7^{+0.7}_{-0.5}\MG} by propagating the errors in the input through the TA model. On top of this we must take into account the intrinsic `cosmic scatter' which we can calculate from the r.m.s. scatter when we apply the TA to our data set (see Table \ref{scatter_table}). Applying this error on top of the upper and lower limits from the TA gives a LG mass of \m{4.7^{+2.9}_{-1.8}\MG}. 
Using the velocity observations with Sal. 2015, we have an estimate with propagated errors of \m{3.8^{+1.1}_{-0.9}\MG}, on top of which we have cosmic scatter of \m{3.8^{+4.7}_{-2.1}\MG}. 

The primary reason for evaluating the TA$_\Lambda$ mass estimate is to set a benchmark for the other methods, although it remains a significant estimator in the literature. The result we retrieve is slightly higher than TA mass in \cite{vdMarel}, as we have included a $\Lambda$ term. 

\subsection{ANN estimates}
To calculate the estimates for each case, we use the ANN which we trained previously with cuts on transverse velocity. For the high transverse velocity case we may use \m{v_t \le 250 \text{km s}^{-1}}, and for the low transverse velocity case we may use \m{v_t \le 62.5 \text{km s}^{-1}}. One could approach this by making stricter cuts on each of the parameters to create a LG like set, but we do not have a large enough set to collect enough pairs for the training, validation, and testing sets in such a narrow band of parameters. Errors in the input parameters are propagated through by running the ANN with all variations of mean, max, and min parameters for \m{(r,v_r,v_t)}. 

\subsubsection{Using only $(r,v)$}
Using vdM. 2012, which has the more conventional (smaller) transverse velocity, we have an estimate of \m{3.6^{+0.3}_{-0.3}\MG} by propagating errors from the input. This error is dwarfed however by the random scatter, which gives us \m{3.6^{+1.9}_{-1.3}\MG}.
With the higher transverse velocity and lower radial velocity of Sal. 2015 we get an estimate of \m{3.3^{+0.6}_{-0.5}\MG} by propagating the errors through. Once again, this is small compared the intrinsic scatter, \m{3.3^{+2.3}_{-1.4}\MG}. The very small variation due to propagating the uncertainties in the inputs is a consequence of the clustering around the mean mass; the ANN trained on $(r,v_r)$ is strikingly insensitive to input.  

\subsubsection{Including shear}
The estimate of the LG mass is boosted when shear information is taken into account. Using this ANN model, we obtain estimates of ${4.9^{+0.8+1.7}_{-0.8-1.3}}$ and $3.6^{+1.3+2.1}_{-1.1-1.3}$ for vdM. 2012 and Sal. 2015 respectively (propagated errors are quoted first, followed by intrinsic scatter).There is a somewhat larger propagated error from the uncertainties in the input than even the TA$_\Lambda$; this is both due to having more parameters in our inputs and the difficulty in acquiring very accurate shear information. The velocity shear tensor identifies the MW-M31 system as being in a filament; the largest collapse eigenvector is almost perpendicular to the radial separation vector (and hence can be safely ignored), but there is an appreciable component along the other two. Although one of the eigenvalues indicates collapse, we also have a large expansion term with an appreciable component along the radial separation direction. This may be causing the overall boost in the system mass when included in the calculation. The effect of the shear compared to the TA is well within the errors of either method, so it is not necessarily possible to say what additional dynamics are at play, or whether the mass estimate is distinctly higher in the presence of shear for the case of the LG. 

\subsection{A probabilistic approach}

\begin{figure} 
  \caption{Normalised histograms of the simulation samples, with each estimate of transverse velocity. There are 118 galaxy pairs with $v_t \le 34$ km s$^{-1}$, and 349 with $57$ km s$^{-1} \le v_t \le 242$ km s$^{-1}$.}
   \label{Mass_Sample}
\includegraphics[width=8cm,keepaspectratio=true]{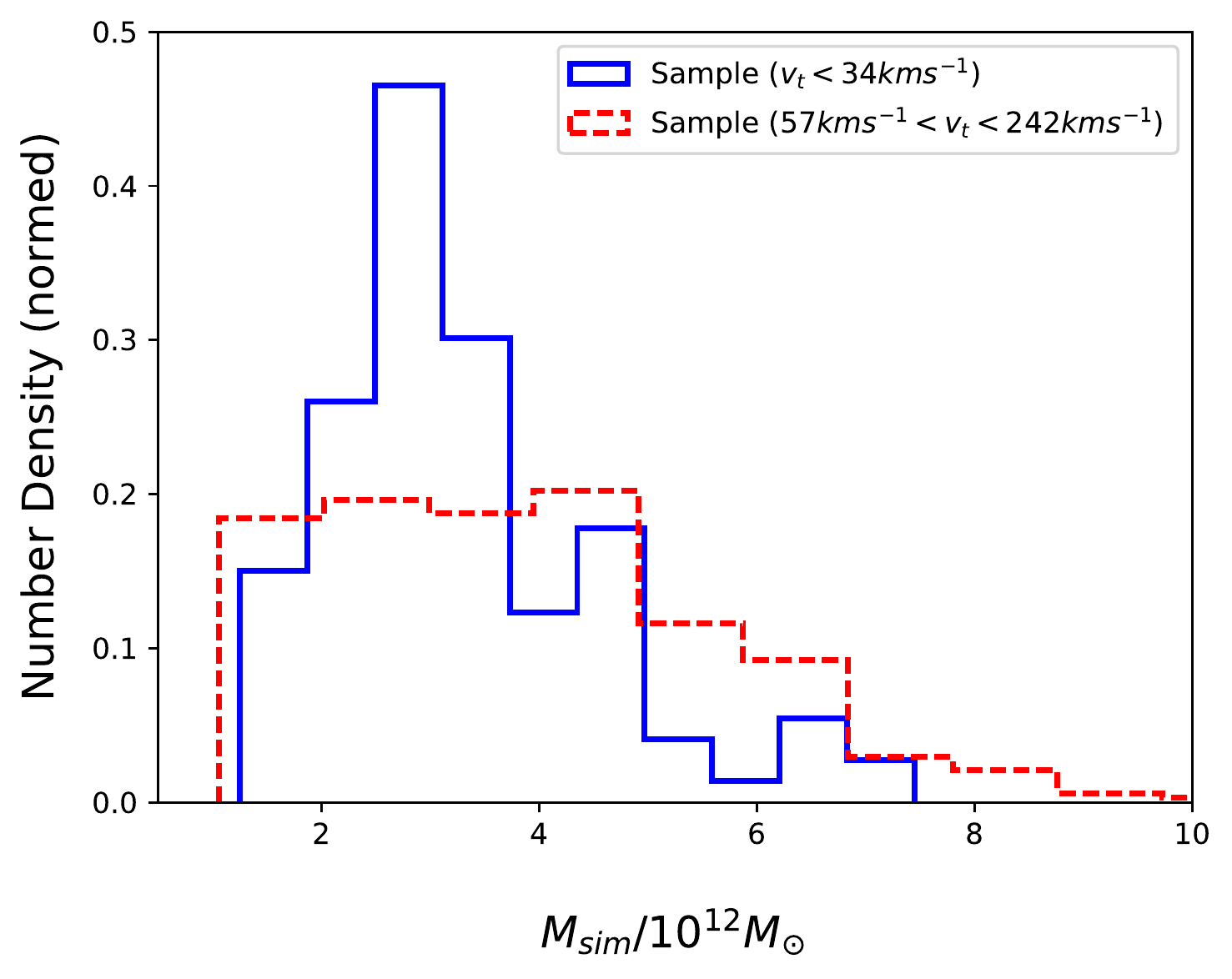}
\end{figure}

Another approach is to look at the simulation data from a Bayesian perspective to estimate a probability distribution for mass, given the masses in the simulations. This can be calculated using an MCMC approach or a numerical integration. A simple approach is to start by cutting the data to select only binaries which have parameters within the limits set by observations of $r$, $v_r$, and $v_t$ for the LG. A probability distribution for the simulation masses can then be fitted to this subset of the data, which is assumed to be Gaussian in the log mass with some mean \m{\mu} and standard deviation \m{\sigma}. A probability distribution over the mass given the ensemble of masses from the simulation is given by the following expression:
\begin{equation}\label{P_M}
	P(M|\textbf{M}_\text{sim}) = \int P(M|\mu,\sigma ,\textbf{M}_\text{sim}) P(\mu,\sigma|\textbf{M}_\text{sim}) d\mu d\sigma
\end{equation}
where \m{\textbf{M}_\text{sim}} is the data vector, composed of each binary $i$ with mass $M_\text{sim}^i$ in the cut of data (these data vectors are represented as histograms in Figure \ref{Mass_Sample}), and \m{\mu,\sigma} characterise the probability distribution in the log mass. Since we don't know the mean and standard deviation of the underlying probability distribution a priori, we are marginalising over values of \m{\mu} and \m{\sigma} in equation \ref{P_M}. We must then calculate the probability distribution on \m{(\mu,\sigma)} as follows:
\begin{equation}
	P(\mu,\sigma|\textbf{M}_\text{sim}) = \frac{P(\textbf{M}_\text{sim}|\mu,\sigma)P(\mu,\sigma)}{P(\textbf{M}_\text{sim})}
\end{equation}
We must now make some assumptions in order to proceed. As mentioned above, we will assume that masses in $\textbf{M}_\text{sim}$, as well as the LG mass, are drawn from an underlying Gaussian distribution in the log mass, and that samples are independent:
\begin{equation}
	P(\log(M)|\mu,\sigma) =  \frac{1}{\sigma \sqrt{2\pi}} \exp \left( \frac{(\log(M)-\mu)^2}{2\sigma^2} \right)
\end{equation}
\begin{equation}
	P(\textbf{M}_\text{sim}|\mu,\sigma) = \prod_{i} P(\log(M_{\text{sim}}^i) |\mu,\sigma)
\end{equation}
We shall also assume flat priors (since we have no prior knowledge) on the parameters \m{\mu} and \m{\sigma}, which means that both \m{P(\mu,\sigma)} and \m{P(\textbf{M}_\text{sim})} are simply multiplicative constants. (In practice we need to assume some hard limits on \m{\mu} and \m{\sigma} so that we can terminate the integration, so their distribution is a top hat function which determines the integration limits for equation \ref{P_M}. The limits must be wide enough so that \m{P(\textbf{M}_\text{sim}|\mu,\sigma)} is negligible outside; we have taken $\mu \in [0,1]$ and $\sigma \in [0.1,1]$.) 

\begin{figure} 
  \caption{Normalised posterior distributions for the mass of the LG calculated from equation 15, with each estimate of transverse velocity.}
   \label{Mass_Prob}
\includegraphics[width=8cm,keepaspectratio=true]{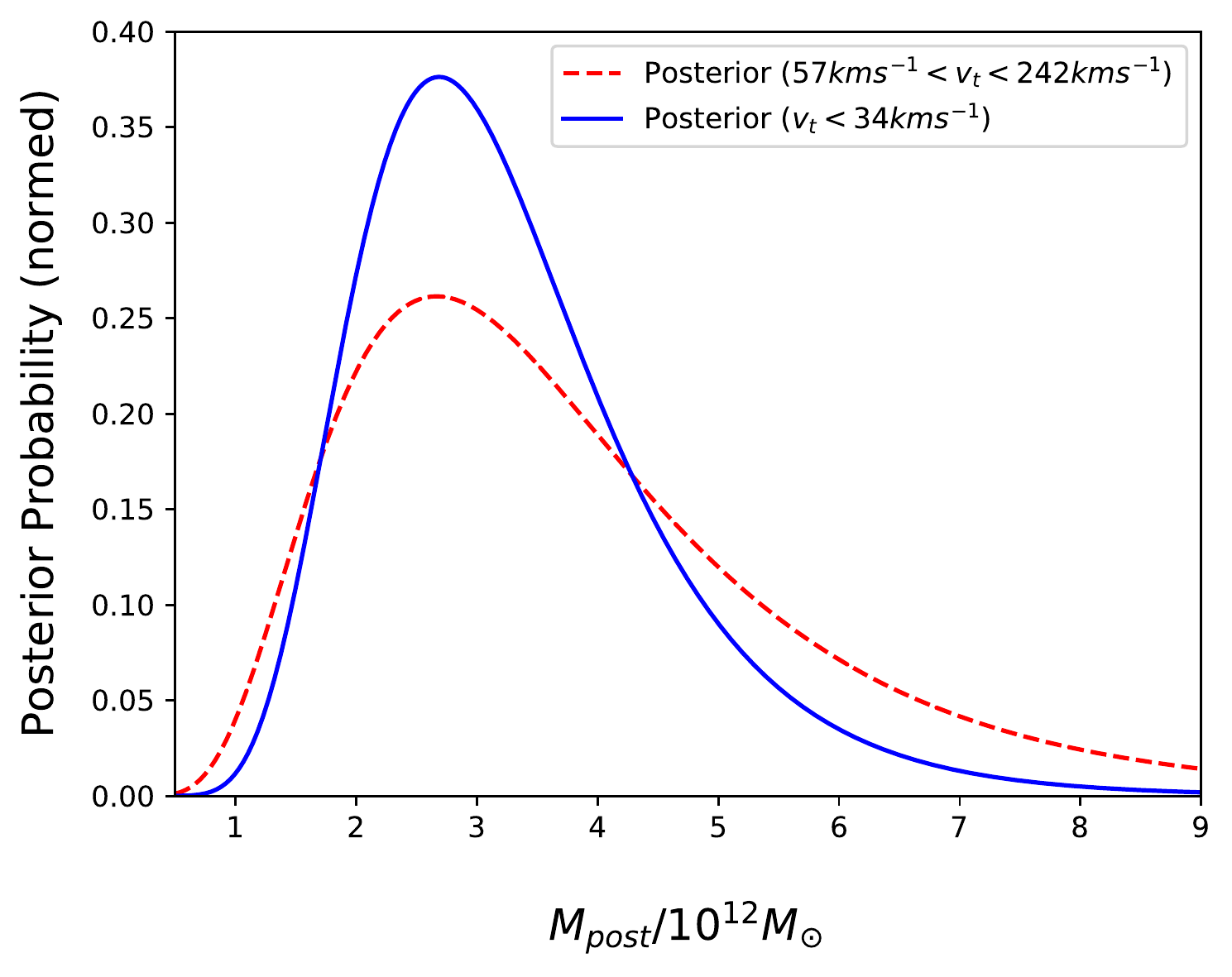}
\end{figure}

With these equations it is simple to evaluate \m{P(M|\textbf{M}_\text{sim})} for a particular cut (see Figure \ref{Mass_Prob}). The posterior resembles to some extent the histograms in Figure \ref{Mass_Sample}, but the Bayesian approach allows us to take into account uncertainties -- particularly those associated with the small number of samples -- in a more robust way, as well as allowing confidence intervals to be calculated. Cuts are made to match the error bars given at the beginning of the section with the exception of \m{v_r} which ranges between 0.5 and 1.5 times the best estimate of $v_r$ in each case in order to give use enough pairs. (A strict cut on all parameters yields less than two dozen pairs for each set.) For a system with low transverse velocity  we find a mass of \m{3.1^{+1.3}_{-1.0}\MG}, and for high transverse velocity we find a mass of \m{3.2^{+2.1}_{-1.2}\MG} at the 68\% confidence level. Best estimates are calculated by the median because this is the same point whether we use a distribution in M or log(M), whereas the peak moves. The median is the peak of the distribution in the log, which is most comparable to the other results since they were calculated using in the log mass. This is largely compatible with the results from ANN, although errors have been calculated in a slightly different way so they may not be directly comparable.

The disadvantages of this method for generating a precise estimate are two-fold. Firstly, we do not have enough pairs to sufficiently cut the data to match observations whilst leaving sufficient pairs to make a statistical judgement calculable. (Even with the relaxed cuts we have made, there are only $\mathcal{O}(10^2)$ pairs, which is not enough to make a strong statistical statement on such a varied group.) Secondly, it does not produce any information on the dependence on different parameters. In order to do this, we would have to create a parameterised probability distribution $P(r,v_r,...)$ and solve for it. It is challenging to produce a flexible enough function form for this distribution in terms of multiple inputs without essentially entering the world of ML. (Methods such as ANN can, after all, be cast as a maximal likelihood method.) Rather, this procedure may be taken as illustrative, and providing a rough probability distribution of LG-like galaxy masses. It demonstrates that the distribution is broad, particularly when the transverse velocity is higher, and the varied results in the literature (see Table \ref{Compare_table}) are well within a sensible range of this distribution. 

\section{Discussion}

\begin{table}
\centering
\caption{A table comparing LG mass estimates of some recent papers. Errors between papers have been estimated with various methods and confidences, and may not be directly comparable. Results are compiled from Li \& White (2008) \cite{LW}; Phelps, Nusser, \& Desjacques (2013) \cite{Phelps}; Pe\~narrubia et al. (2015) \cite{Penarrubia}; Gonz\'alez, Kravtsov, \& Gnedin (2014) \cite{Gonzalez}; van der Marel (2012) \cite{vdMarel}; Carlesi et al. (2016) \cite{Carlesi}.}\label{table:Mass_Table}
\label{Compare_table}
\renewcommand{\arraystretch}{1.5}
\begin{tabular}{|c|c|c|c|c|}
\hline
Paper  & Method & Mass & Comments / other Esimates \\ 
\hline
\cite{LW}   & Large $\Lambda$CDM & $5.27^{+4.96}_{-1.82} \MG$ & $r = 784$ kpc  $v_r = 130$ km s$^{-1}$\\
&   n-body Sim. & & $M_{TA} = 5.32 \pm 0.48 \MG$ \\
\hline
\cite{Phelps}  & Numerical Action & $6\pm1  \MG$ & $ r = 790 $ kpc, $v_r = 119$ km s$^{-1}$\\
&  Method & & $M_\text{MW} = 2.5\pm 1.5 \MG$   \\
& & & $M_\text{M31} = 3.5 \pm 1 \MG$ \\
\hline
\cite{Penarrubia}  & Extended LG & $2.3\pm0.7 \MG$ & $r = 783$ kpc, $v_h = 300$ km s$^{-1}$ \\
 &  Simulation & & Uses heliocentric coordinates \\
\hline
\cite{Gonzalez}  & Large $\Lambda$CDM & $4.2^{+3.4}_{-2.0} \MG$ & $r = 770$ kpc, $v_r = 109.3$ km s$^{-1}$ \\
&  n-body Sim. & & Finds $M_{MW} + M_{M31} = 2.4 \MG$\\
\hline 
\cite{vdMarel}  & TA ($\Lambda = 0$) & $4.93 \pm 1.63 \MG$ & $r = 770$ kpc, $v_r = 109.3$ km s$^{-1}$ \\
& converted to virial & &  $M_{TA} = 4.27\MG$ \\
& & & Multi-probe $M = 3.17\MG$ \\
\hline
\cite{Carlesi} & Constrained & $1.79 \pm 0.51 \MG$ & $0.35 $ Mpc $ < r < 0.7$ Mpc\\
& Simulations & & $-135 $ km s$^{-1} < v_r < -80$ km s$^{-1}$ \\
\hline
\end{tabular}
\end{table}

We find that, despite some advance in the analytic models, the application of ML provides a rapid route to improved estimates of the LG mass. Analytically, the inclusion of \m{\Lambda} in the TA can provide an important improvement in the bias of the mass estimate when analysing a large collection of galaxies from $\Lambda$CDM simulations, however the inclusion of $v_t$ is if anything disruptive. The uncertainty in \emph{which} parameters to base our model around is a significant stumbling block for this calculation, and ML methods are a valuable tool in analysing the potential of different parameters to improve the solution. Furthermore we find that the application of ANN presents a more significant improvement than analytic extensions to the TA. When trained (on a suitably large population) using only parameters \m{(r,v_r)} one finds the scatter in the log mass which is reduced by approximately half, but the estimator highly biased towards the mean mass of the sample, so that it fails to make predictions. Additional parameters, particularly the shear field, pushes this scatter lower, approximately half of the TA scatter and with a significantly reduced bias compared to the ANN with only $(r,v_r)$ when tested on a sample with no cuts to $v_t$. 

The velocity shear tensor provides a key improvement to the ANN prediction of mass. Without it, the ANN cannot find an explanation for why many halo pairs which appear to have high dynamical masses (judging from the TA) actually have much lower masses in the simulations; this leads to a mass cap for the ANN function in order to reduce the scatter. The addition of shear information breaks to some extent this uncertainty and produces an improved correlation between $\log(M)_{\text{ANN}}$ and $\log(M_{\text{sim}})$. Since the influence of this parameter is poorly understood from an analytic perspective, an investigation into the behaviour of the derived function from the ANN trained on shear information may be used to gain physical insight. A possible explanation is that there is a change in mass estimate due to the bulk motion on large scales affecting the motion of the binary system. In other words, our halo pairs cannot be taken to interact in isolation, but their place within in the cosmic web structure determines their motion to some extent due to local velocity flows creating `expansion' or `collapse' along different axes. This is an effect not previously included in studies of the LG mass, and hints at the deficiencies of our isolated physical models. 

Applications of trained ANN using the observed LG parameters provide an estimate of the mass of the MW and Andromeda (see Table \ref{LG_table}). This is consistent with other mass estimates, of which some recent examples are summarised in Table \ref{table:Mass_Table}. Some mass estimates of the LG are also summarised in \cite{Courteau}, which tend to range from 2.5 -- \m{5 \times 10^{12} M_{\odot}}. There is some tension in the results of the LG mass, with a divide between low estimates \cite{Penarrubia} \cite{Carlesi}, and high estimates \cite{LW}\cite{Partridge}\cite{Phelps}\cite{vdMarel}. One potential source of discrepancies is the assumption that $M_\text{LG} \approx M_\text{MW} + M_\text{M31}$. In \cite{Gonzalez}, they find a significant mass difference between the LG ($4.2\MG$) and the sum of the two major galaxies ($2.4 \MG$). Part of this may be due to the fact that dynamical methods that work on the basis of a binary system implicitly include the mass of any bound satellites with the mass of the galaxy. Hence, the Milky Way or Andromeda mass found by such a method would include all the satellites that move with it, and thus push the mass estimates of the individual galaxies much higher. In \cite{Carlesi} they find a particularly low LG mass, at only $1.8\MG$. It is notable however that the range of $r$ is much lower than in the other studies, being contained in the range $0.35$ Mpc $< r < 0.7$ Mpc, whereas others (including this one) use $r \sim 0.77$ Mpc. This may go some way towards explaining the comparably low result. For example, taking a TA estimate for the LG mass from the middle of their range, which would have $r = 0.525$ Mpc, and $v_r = 107.5$ km s$^{-1}$, gives $M_{TA_{\Lambda}} = 2.3_{-1.1}^{+1.6} \MG$, with the lower and upper bounds calculated using $r = 0.35$ Mpc and $0.7$ Mpc respectively. The mean (using $r = 0.525$ Mpc) is just at the upper estimate of \cite{Carlesi} ($1.8 \pm 0.5 \MG$), and with significant overlap in errors in the TA are taken into account. This suggests that the results in \cite{Carlesi} may indeed be consistent with this paper (and others in line with the TA) after all. 

Given that there is only one LG, statistical methods cannot overcome the possibility that the LG is a `special case', occupying an less than maximally likely part of the distribution, due to some hitherto unknown or unobservable parameters. The action of additional bodies, as in \cite{Penarrubia} and \cite{Phelps}, may have an effect as well, which could make the system highly sensitive to configurations. It is also suggested in \cite{Penarrubia} that the uncertainties in the sun's transverse motion may be a significant source of error in determining correctly the relative motions; it also however asserts that the TA must be systematically biased by a factor of about 1.5 in order to match with their result, which is not seen in this study or in \cite{LW} for galaxies with small transverse velocities. All of these mass estimation methods have significant error bars, and overlap at the one or two sigma level. Refinement of these methods will be crucial to identify a genuine tension between these results, which could point to the need to new physical models.

When we apply the ANN trained on shear information, we have a mass estimate of \m{4.9^{+0.8+1.7}_{-0.8-1.3}\MG}, compared to a TA\m{_\Lambda} estimate of \m{4.7^{+0.7+2.9}_{-0.6-1.8}\MG}. These are very similar, although the scatter is significantly improved. Bayesian methods based on a cut of data from the simulation also offer a competitive estimate of \m{3.1^{+1.3}_{-1.0}\MG}, although this is forced to be use few parameters and thus effectively averages over any other influences such as shear effects. (Making strict cuts on the all the parameters, including shear parameters, would require a larger dataset than used in this paper; a Bayesian analysis based on cuts on all parameters would require millions of halo pairs.) It is therefore unsurprising that the Bayesian analysis produces a result similar to the ANN trained on just two parameters (see Table \ref{LG_table}). It is interesting to compare the ANN results with the Bayesian analysis on the data; if we look at the distributions in Figure \ref{Mass_Prob} we notice that the value estimated by the shear ANN is not by any means surprisingly unlikely. The distributions in Figure \ref{Mass_Prob} really represent the distributions when shear effects are averaged over, and hence the probability of the ANN mass estimate in the context of the marginalised probability distribution is related to the probability of being in an LG-like local shear environment in the simulation. 

To improve the estimates for the LG, one may wish to refine this estimate by using a larger simulation, selecting more strictly ``LG like" pairs, and using additional criteria such as morphology as in \cite{LW}. Applications to the LG may also be complicated by uncertainties in additional parameters. The transverse velocity (usually taken to be small) has been contested recently, implying that strict cuts on \m{v_t} in LG-like sets may not be justified, which jeopardises the use of simple dynamical models such as the TA. In such a case it is particularly useful to have methods such as ML; the ANN estimates have only slightly higher scatter for analyses with no cut to $v_t$ as for sets restricted to low $v_t$. The results of both the exploration of the TA with a Cosmological Constant term (TA\m{_\Lambda}) and the significance of the shear tensor (\m{\Sigma_{ij}}) indicate that successful dynamical estimates of mass are unlikely to be independent of cosmological context. We note that taking into account both the cosmological model (in order to model expansion terms such as \m{\Lambda} or more complex models such as modified gravity or dark energy) and the larger scale structure (as traced by \m{\Sigma_{ij}}) is crucial to understanding how galaxies are interacting. Our methods are dependent upon $\Lambda$CDM simulations, and thus are only applicable within this framework. Estimates of the LG mass in the context of other cosmological models -- alternative dark energy models, modified gravities, or warm dark matter -- could yield different results and must be explored separately using appropriate simulations. It is possible that a greater understanding of galaxy formation physics will inform such mass estimates even better, providing clues to possible boundary conditions for dynamical arguments, or linking galaxy masses to their environments and cosmological models in a more direct way. 

\section*{Acknowledgements}

The authors would like to thank David Klein, Donald Lynden-Bell, and Candace Partridge for their prior work and illuminating discussions on this topic. We would also like to thank Jaime Forero-Romero for providing us with simulation data used in preliminary work for this paper. 

MM is supported by an UCL studentship. OL acknowledges support from a European Research Council Advanced Grant FP7/291329.

The CosmoSim database used in this paper is a service by the Leibniz-Institute for Astrophysics Potsdam (AIP).
The MultiDark database was developed in cooperation with the Spanish MultiDark Consolider Project CSD2009-00064. 

The authors gratefully acknowledge the Gauss Centre for Supercomputing e.V. (www.gauss-centre.eu) and the Partnership for Advanced Supercomputing in Europe (PRACE, www.prace-ri.eu) for funding the MultiDark simulation project by providing computing time on the GCS Supercomputer SuperMUC at Leibniz Supercomputing Centre (LRZ, www.lrz.de). The Bolshoi simulations have been performed within the Bolshoi project of the University of California High-Performance AstroComputing Center (UC-HiPACC) and were run at the NASA Ames Research Center. Simulation data was retrieved from the cosmosim.org database.

\end{document}